\documentstyle[twoside,fleqn,espcrc2,epsf]{article}
\def\fig#1#2#3{\epsfxsize=#3truein
\vskip -0.3 truein
\centerline{\epsffile{fig#1.eps}}
\centerline{\vbox{{\bf \noindent Figure #1.} #2}}
\bigskip}

\def\figsize{2.2}
\def\pbp{\langle\overline{\psi}\psi\rangle}
\def\w{\langle w \rangle}
\def\meff{m_{\rm eff}}

\def\spose#1{\hbox to 0pt{#1\hss}}
\def\ltapprox{\mathrel{\spose{\lower 3pt\hbox{$\mathchar"218$}}
 \raise 2.0pt\hbox{$\mathchar"13C$}}}
\def\gtapprox{\mathrel{\spose{\lower 3pt\hbox{$\mathchar"218$}}
 \raise 2.0pt\hbox{$\mathchar"13E$}}}
\def\inapprox{\mathrel{\spose{\lower 3pt\hbox{$\mathchar"218$}}
 \raise 2.0pt\hbox{$\mathchar"232$}}}

\def\one{Chiral condensate vs. fermion mass.}
\def\two{t' Hooft vertex vs. fermion mass.}
\def\three{Chiral condensate vs. $L_s$ for $m_f=0$.}
\def\four{Chiral condensate vs. $L_s$ for $m_f \neq 0$.}
\def\five{t' Hooft vertex vs. $L_s$ for $m_f \neq 0$.}
\newcommand{\AmS}{{\protect\the\textfont2
  A\kern-.1667em\lower.5ex\hbox{M}\kern-.125emS}}

\title{Domain Wall Fermions and chiral symmetry restoration rate.}

\author{P. Vranas\address{Dept. of Physics, Columbia University,
	New York, NY 10027}
	\thanks{Presented by G. Fleming. This work was supported 
by DOE grant \# DE-FG02-92ER40699. }}
       
\begin{document}

\begin{abstract}

Domain Wall Fermions utilize an extra space time dimension to provide
a method for restoring the regularization induced chiral symmetry
breaking in lattice vector gauge theories even at finite lattice
spacing. The breaking is restored at an exponential rate as the size
of the extra dimension increases. As a precursor to lattice QCD
studies the dependence of the restoration rate to the other parameters
of the theory and, in particular, the lattice spacing is investigated
in the context of the two flavor lattice Schwinger model.

\end{abstract}

% typeset front matter (including abstract)
\maketitle

A massive vector theory in $2n+1$ dimensions with a mass term $m(s)$
that depends on the periodic $2n+1$ dimension $s$ such that $m(s) =
m_0$ for $1 \leq s \leq L_s/2$ and $m(s) = -m_0$ for $L_s/2 < s \leq
L_s$, where $L_s$ is the size of the $2n+1$ dimension, develops a
Dirac fermion in 2n dimensions with chiral components separated and
localized along the $2n+1$ direction $s$ at $s=0$ and $s= L_s/2$
\cite{Kaplan}. Perturbative calculations indicate exponential
localization along $s$ with the mixing of the chiral components
vanishing exponentially fast with $L_s$. At $L_s=\infty$ the mixing is
zero and the regularization leaves intact the chiral symmetry.

The $L_s=\infty$ case can be handled using the Overlap formalism
\cite{NN1}.  Therefore the Overlap is an ideal regulator for vector
theories.  However, calculations using the Overlap involve
diagonalization and determinant evaluation of matrices with size
$\sim$ volume $\times$ volume. Because of this large scale dynamical
QCD calculations with the Overlap are beyond the capacity of present
day supercomputers (however, see \cite{Neuberger} for a promising new
development).

An alternative is to keep $L_s$ finite and view it as an extra
parameter that controls the regularization induced chiral symmetry
breaking. This formulation goes under the name of Domain Wall Fermions
(DWF). Unlike Wilson or staggered fermions the chiral limit can be
approached even at finite lattice spacing by increasing $L_s$. The
theory can be simulated using standard Hybrid Monte Carlo
techniques. However, before DWF can be used in dynamical simulations
of QCD the following questions should be answered:

\noindent
1) Is the regularization induced chiral symmetry breaking restored
exponentially fast as $L_s$ increases in the full dynamical theory?

\noindent
2) How does the functional behavior of the restoration and the
restoration rate depend on the lattice spacing $a$, the fermion mass
$m_f$ and the domain wall height $m_0$?

\noindent
3) For the interesting $a$ and $m_f$ how large should $L_s$ be
for the effects of the regularization induced chiral symmetry breaking
to be negligible?

This work addresses these questions in the context of the $N_f = 2$
flavor lattice Schwinger model at fixed physical volume. For a
detailed analysis see \cite{Vranas}. For some interesting work
that also partially addresses these questions see \cite{Blum-Soni}.  A
variant domain wall model is used \cite{Boyanovsky}\cite{Shamir}
where $m(s) = m_0$ for $1 \leq s \leq L_s$ but the boundary
conditions along the $2n+1$ direction are free. In this model the
chiral components are localized at $s=1$ and $s=L_s$ and as a result
the decay region is twice as long for the same $L_s$. Also, an
explicit coupling of the two chiral components at $s=0$ and $s=L_s$
with strength $m_f$ is introduced to allow for linear control over the
fermion mass.  The action is the same as the one for a $2n+1$ Wilson type
fermion except that the gauge field ``lives'' only on the $2n$
dimensional lattice, the diagonal term is of the form $2n+1-m_0$ and
the boundary conditions along the $2n+1$ dimension are free.

An explicit calculation \cite{Vranas} of the free theory propagator
gives the mass of the lightest mode:
\begin{eqnarray*}
\meff = m_0 (2 - m_0) \left[ m_f + (1-m_0)^{L_s}\right]
\label{meff_free}
\end{eqnarray*}
This strongly suggests the pattern for chiral symmetry restoration 
in the model. Recent calculations of the one loop correction
result in a renormalization of $m_0$ \cite{Aoki}.

Two observables are used to probe the symmetries of the theory:
the chiral condensate $\pbp$
and the  t'Hooft vertex $\w$ where
$
w = \prod_{i=1}^{N_f} \bar\psi^i_R \psi^i_L + \prod_{i=1}^{N_f}
\bar\psi^i_L \psi^i_R
$.
The chiral condensate is used to probe chiral symmetry.  At $m_f = 0$
and $L_s=\infty$ $\pbp = 0$. When $L_s$ is finite the regularization
breaks chiral symmetry and $\pbp \neq 0$. From perturbative
calculations one expects that $\pbp$ should approach zero
exponentially fast with increasing $L_s$.  The t' Hooft vertex is used
to probe the anomalously broken $U(1)$ axial symmetry. At $m_f=0$,
$L_s=\infty$ one should find that $\w \neq 0$.

In order to study the approach to the $L_s \rightarrow \infty $ limit
it is very useful to calculate the values of these observables at $L_s
= \infty$ using the Overlap formalism. In two dimensions this is not
computationally demanding and can be done as in \cite{NNV}.  The
results, interesting in their own right, are shown in figures 1 and 2.
In figure 1 $\pbp / m_\gamma$ is plotted vs. $m_f$ for $L=6$, $m_0
=0.9$ and $\mu l = 3.0$. $g_0$ is the coupling constant $\mu l =
g_0 L / \sqrt{\pi}$, $m_\gamma = \mu \sqrt{N_f}$ is the photon mass,
$L$ is the lattice size in lattice units and $l$ in physical units.
The fits are to $\pbp = A m_f^p$.  Both fits have a $\chi^2$ per
degree of freedom of about one.  For $m_f < 0.1$ $p = 0.996(3)$, while
for $m_f>0.1$ $p = 0.32 (2)$. This is in agreement with the
analytical solutions found in \cite{Smilga-Hetrick}.  In figure 2
$<w>/m_\gamma^2$ is plotted vs. $m_f$ for the same parameters.  The
dotted lines are the $m_f=0$ result $\pm$ the statistical error. As
expected at $m_f=0$, $\w \neq 0$.

The finite $L_s$ model is studied using a standard Hybrid Monte Carlo
algorithm. In order to reveal the chiral symmetry breaking due to the
DWF regularization, $\pbp$ is calculated for various $L_s$ at
$m_f=0$. It is plotted in figure 3 for fixed physical volume $\mu l =
3.0$, $m_0 = 0.9$ and four lattice spacings $\mu l / L = \mu a$, with
$L= 6, 8, 10, 12$ corresponding to the lines from top to bottom. The
data is consistent with exponential decay indicated by the fitted
lines ($\chi^2$ per degree of freedom $\approx 1$).  However, at the
larger spacing $L=6$ the data also fits well to a power law
behavior. This is not so at the smallest spacing $L=12$ (a power law
fit has $\chi^2$ per degree of freedom $\approx 31$).

\fig{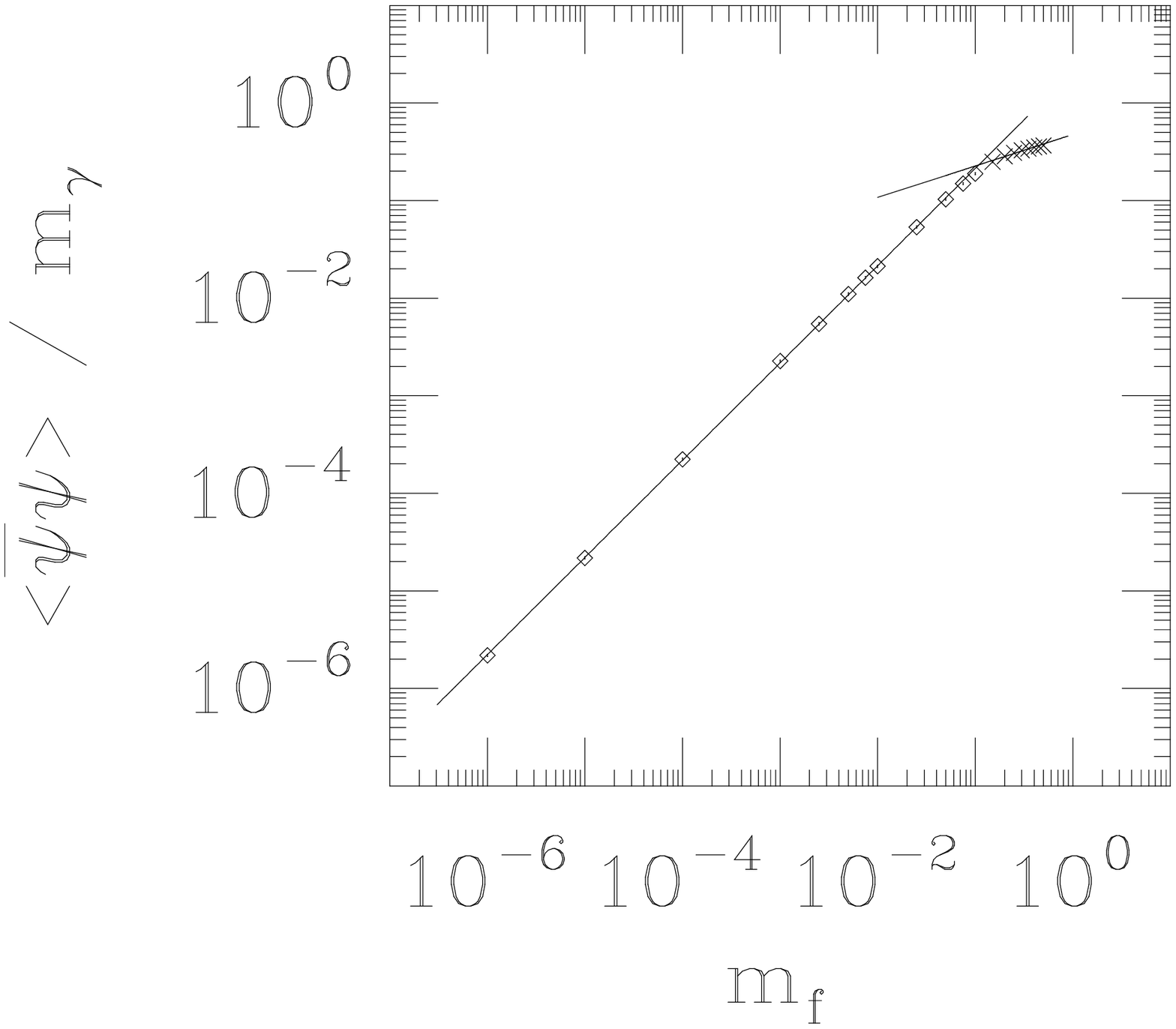}{\one}{\figsize}

\fig{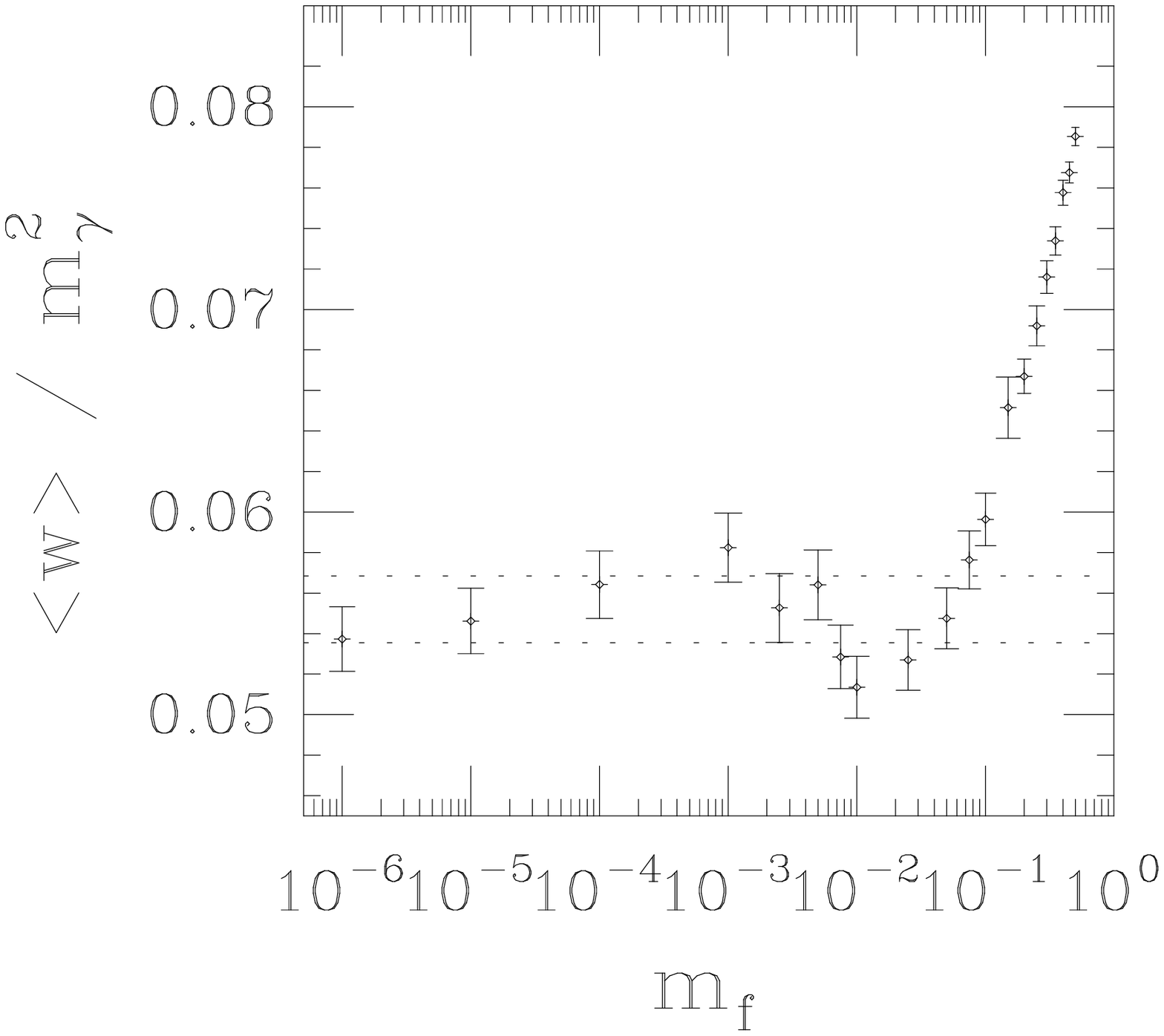}{\two}{\figsize}

\fig{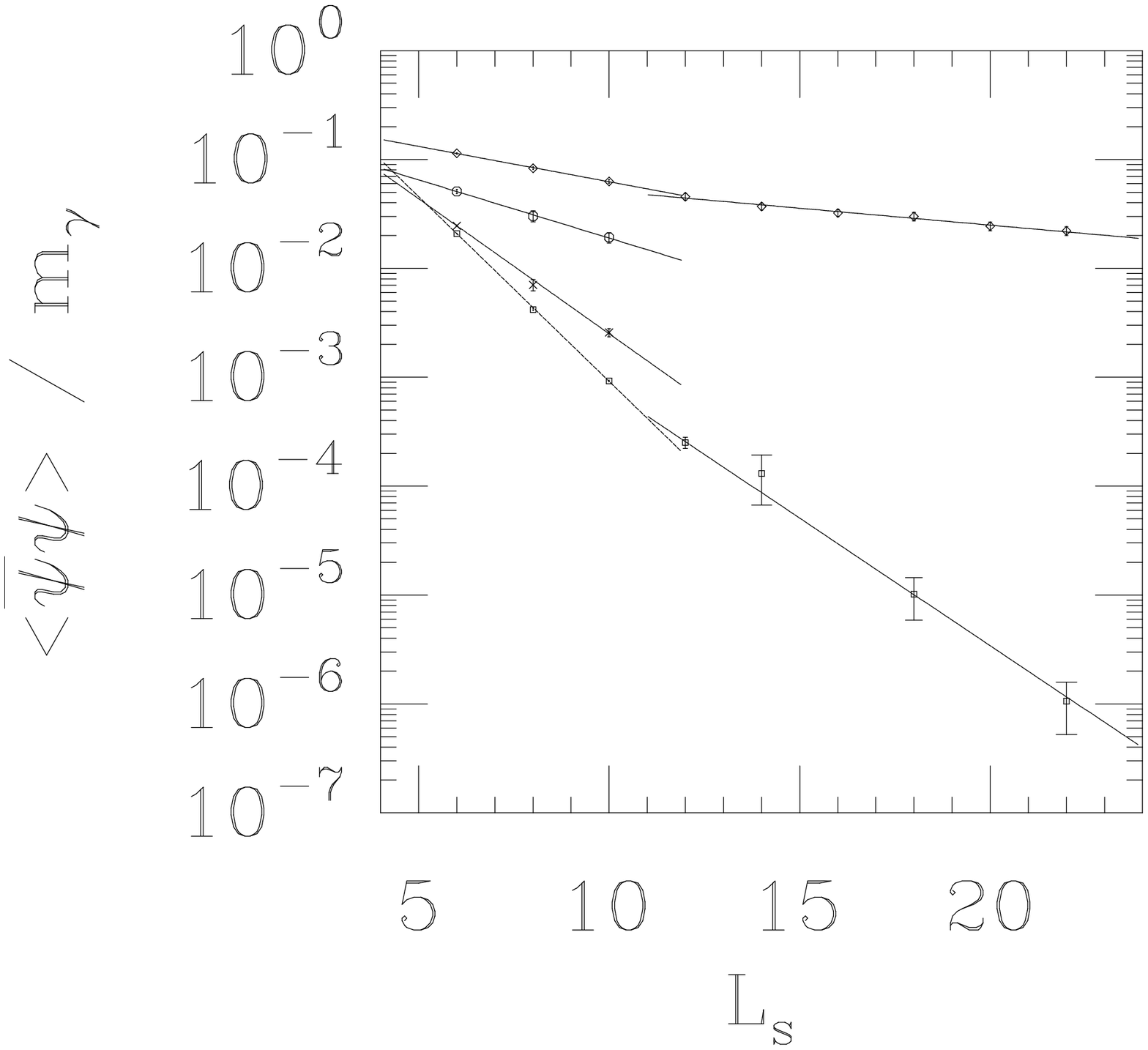}{\three}{\figsize}

This behavior is consistent with a picture where chiral symmetry is
restored with a fast exponential decay rate for $L_s$ up to some value
and with a slower exponential decay rate for $L_s$ above that
value. For the range of lattice spacings used, the inflection appeared
at $L_s \approx 10$.  Using a simple model, it was found in
\cite{Vranas} that the first fast decay can be associated with
restoration of chiral symmetry in the zero topological sector while
the second slower decay can be associated with the regions of gauge
field configuration space that connect the $q=0$ and $q=\pm 1$
topological sectors.

\fig{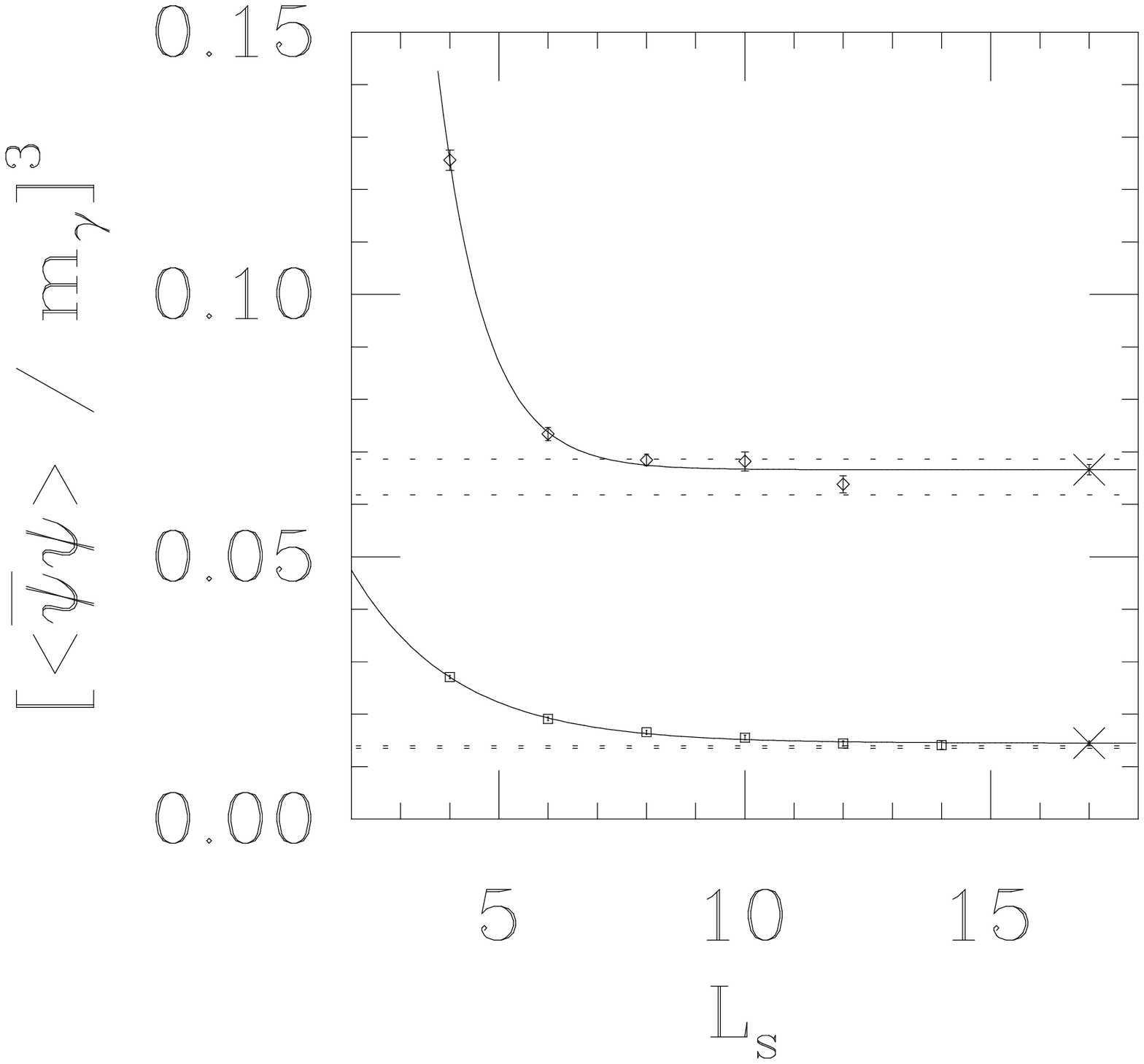}{\four}{\figsize}

\fig{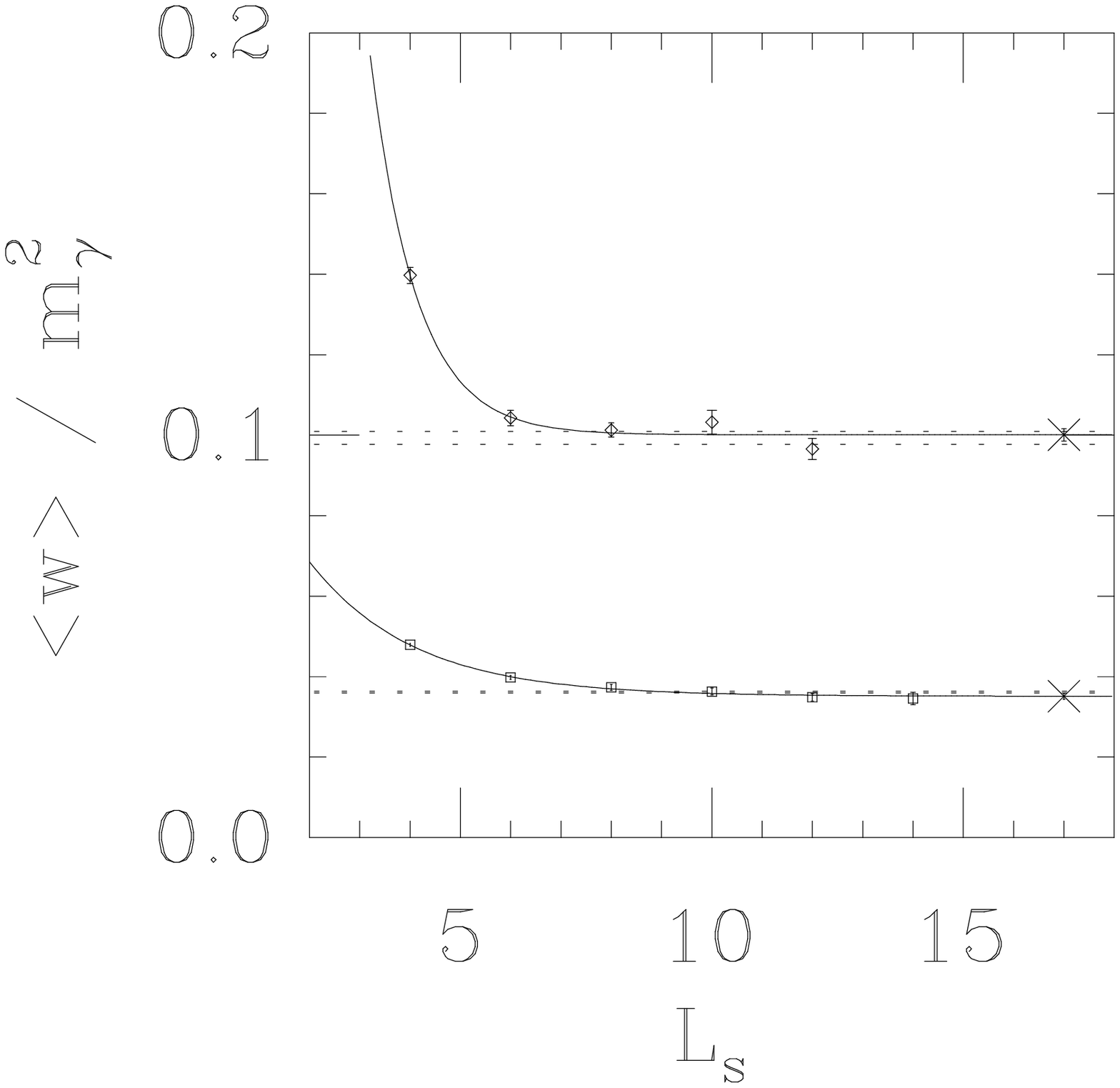}{\five}{\figsize}

The effects of the size of the lattice spacing $a$ to the two decays
are apparent. The fast decay becomes faster as $a$ is decreased.  The
vanishing of $\pbp$ is consistent with a form $e^{-c L_s}$ with
$e^{-c}$ being roughly a linear function of $a$. However, more data at
smaller $a$ are needed before one can be confident that this is the
correct scaling form.  The second slower decay also becomes faster as
$a$ is decreased and it differs less from the faster decay as $a$
becomes smaller.  This behavior can also be understood using the
simple model in \cite{Vranas}.

For small but non zero $m_f$ the values of $\pbp$ and $\w$ are
presented in figures 4 and 5 for two different lattice spacings set by
$L=4$ (lower curve) and $L=10$ (upper curve) at $m_0=0.9$.  The
physical volume and $m_f L$ are fixed at $\mu l = 3.0$ and $m_f L =
2.0$. The fits are to $A + B e^{-c L_s}$. The dotted lines are the
$L_s=\infty$ results $\pm$ the error.  The cross is the coefficient
$A$.  The $L_s = \infty$ numbers are approached in a way that is
consistent with exponential decay with a rate that becomes faster as
the lattice spacing decreases.  Finally, it was also found in
\cite{Vranas} that the larger the fermion mass the sooner the $L_s =
\infty$ value was approached (within a few percent at $L_s = 4 - 8$).

The next step is to carry out a similar investigation for dynamical
QCD. Many of the characteristics of DWF found here are sufficiently
generic so that one would expect that they will also be present in
QCD. If it turns out that QCD at the presently accessible lattice
spacings, volumes and quark masses has similar restoration rates as
the ones found here, then DWF will indeed provide a powerful fermion
discretization method.

\end{document}